\documentclass{article}
\usepackage{amssymb, amsmath}

\usepackage{graphics}
\usepackage[dvips]{graphicx}
\usepackage{eepic}
\usepackage{epic}
\usepackage{array}
\usepackage{bm}
\usepackage{exscale}
\usepackage{latexsym}

\newtheorem{thm}{Theorem}[section]

\newtheorem{prp}[thm]{Proposition}

\newtheorem{dfn}[thm]{Definition}
\def\C{{\mathbb C}}
\def\Z{{\mathbb Z}}
\def\Q{{\mathbb Q}}
\def\F{{\mathbb F}}
\def\R{{\mathbb R}}
\def\P{{\mathbb P}}
\def\A{{\mathbb A}}
\def\II{${}_{\mbox{\scriptsize{II} }}$}
\def\I{${}_{\mbox{\scriptsize{I} }}$}
\def\sp{\mbox{\scriptsize p}}
\def\hugesymbol#1{\mbox{\strut\rlap{\smash{\Huge$#1$}}\quad}}
\title{Discrete Painlev\'{e} equations and discrete KdV equation over finite fields}
\author{Masataka \textsc{Kanki}\footnote{Graduate School of Mathematical Sciences, University of Tokyo, 3-8-1 Komaba, Tokyo 153-8914, Japan\newline e-mail: \texttt{kanki@ms.u-tokyo.ac.jp}}
          ~and Jun \textsc{Mada}\footnote{College of Industrial Technology, Nihon University, 2-11-1 Shin-ei, Narashino, Chiba 275-8576, Japan\endgraf e-mail: \texttt{mada.jun@nihon-u.ac.jp}}
~ and Tetsuji \textsc{Tokihiro}\footnote{Graduate School of Mathematical Sciences, University of Tokyo, 3-8-1 Komaba, Tokyo 153-8914, Japan
\newline e-mail: \texttt{toki@ms.u-tokyo.ac.jp}}
}

\begin{document}

\maketitle

\begin{abstract}
We investigate some of the discrete Painlev\'{e} equations (dP\II, $q$P\I and $q$P\II) and the discrete KdV equation over finite fields\footnote{This is a review article on the recent developments in the theory of discrete integrable equations over finite fields based on \cite{KMTT,JNMP,KMT}. This article is accepted for publication in \textit{RIMS K\^{o}ky\^{u}roku Bessatsu} in 2013 as the proceedings of the domestic conference `Expansion of the theory of nonlinear discrete integrable systems' in RIMS, Kyoto, Japan, on August 2012.}. The first part concerns the discrete Painlev\'{e} equations. We review some of the ideas introduced in our previous papers \cite{KMTT, JNMP} and give some detailed discussions. We first show that they are well defined by extending the domain according to the theory of the space of initial conditions.
We then extend them to the field of $p$-adic numbers and observe that they have a property that is called an `almost good reduction' of dynamical systems over finite fields.
We can use this property, which can be interpreted as an arithmetic analogue of singularity confinement, to avoid the indeterminacy of the equations over finite fields and to obtain special solutions from those defined originally over fields of characteristic zero.
In the second part we study the discrete KdV equation. We review the discussions in \cite{KMT} and present a way to resolve the indeterminacy of the equation by treating it over a field of rational functions instead of the finite field itself. Explicit forms of soliton solutions and their periods over finite fields are obtained. 
\end{abstract}

\section{Introduction}
In this article, we study several discrete integrable equations over finite fields.
One of the problems we encounter when we treat a discrete equation over a finite field is that its time evolution is not always well defined. 
This problem cannot be solved even if we extend the domain from $\F_{p^m}$ ($p$ is a prime number) to the projective space $\P\F_{p^m}:=\F_{p^m}\cup\{\infty\}$, because $\P\F_{p^m}$ is no longer a field;
we cannot determine the values such as $\frac{0}{0},\ 0 \cdot \infty,\ \infty+\infty$ and so on.
In this paper we study two strategies to define the time evolution over a finite field without inconsistencies: [I]
One is to reduce the domain so that the time evolution will not pass the indeterminate states.
[II] The other is to extend the domain so that it can include all the orbits. 
We take the latter strategy [II] and adopt two approaches:
(i) The first approach is the application of the theory of space of initial conditions developed by Okamoto \cite{Okamoto,Okamoto2,Okamoto3,Okamoto4} and Sakai \cite{Sakai}.
(ii) The second approach is to extend the domain of initial conditions to the field of $p$-adic numbers $\mathbb{Q}_p$.

The first part of this article concerns the one-dimensional dynamical systems, in particular the discrete Painlev\'{e} equations.
Through the approach (i), we show that the dynamics of those equations can be well defined in the space of initial conditions even over the finite fields \cite{JNMP}.
The approach (ii) is closely related to the so called arithmetic dynamics, which concerns the dynamics over arithmetic sets such as $\mathbb{Z}$ or $\mathbb{Q}$ or a number field that is of number theoretic interest \cite{Silverman}.
In arithmetic dynamics, the change of dynamical properties of polynomial or rational mappings give significant information when reducing them modulo prime numbers. 
The mapping is said to have good reduction if, roughly speaking, the reduction commutes with the mapping itself \cite{Silverman}. The action of the projective linear group $\mbox{PGL}_2$ give typical examples of mappings with a good reduction.
The QRT mappings \cite{QRT} and several bi-rational mappings over finite fields have been investigated in terms of integrability by choosing the parameter values so that indeterminate points are avoided
\cite{Roberts,Roberts2,Roberts3}.
They also have a good reduction over finite fields.
We prove that, although the integrable mappings generally do not have a good reduction modulo a prime,
they do have an \textit{almost good reduction}, which is a generalised notion of good reduction.
We first treat in detail the discrete Painlev\'{e} II equation (dP\II ) \cite{RGH} over finite fields, on which we have briefly reported in our previous letter \cite{KMTT}, and then apply the method to the $q$-discrete Painlev\'{e} equations ($q$P\I and $q$P\II ).
The time evolution of the discrete  Painlev\'{e} equations can be well defined generically, via the reduction from a local field $\Q_p$ to a finite field $\F_p$. The theory is then used to obtain some special solutions directly from those over fields of characteristic zero such as $\Q$ or $\R$.

The second part concerns the two-dimensional evolution equation, in particular, the discrete KdV equation (dKdV). Since the dKdV equation evolves as a two-dimensional lattice, the number of singular patterns we obtain is too large to be investigated even for a system with small size. Therefore, it is difficult to determine the space of initial conditions through the approach (i) above.
Instead, we take the second approach (ii). We can well-define the dKdV equation by considering it over the local field $\mathbb{Q}_p$ just like the discrete Painlev\'{e} cases.
We introduce another extension of the space of initial conditions.
Since the dKdV equation has a parameter $\delta$, we can see $\delta$ as an indeterminant (variable) and can define the dKdV equation over the field of rational functions $\mathbb{F}_r(\delta)$ where $r=p^m\ (m\ge 1)$ \cite{KMT}. We define the soliton solutions over the finite fields and discuss their periodicity.

\section{The dP\II equation and its space of initial conditions}\label{sec2}
The discrete Painlev\'{e} equations are non-autonomous, integrable mappings which tend to some continuous Painlev\'{e} equations for appropriate choices of the continuous limit \cite{RGH}.
They are non-autonomous and nonlinear second order ordinary difference equations with several parameters.
When they are defined over a finite field, the dependent variable takes only a finite number of values and their time evolution will attain an indeterminate state in many cases for generic values of the parameters and initial conditions.
For example, the dP\II equation is defined as 
\begin{equation}
u_{n+1}+u_{n-1}=\frac{z_n u_n+a}{1-u_n^2}\quad (n \in \mathbb{Z}),
\label{dP2equation}
\end{equation}
where $z_n=\delta n + z_0$ and $a, \delta, z_0$ are constant parameters \cite{NP}.
Let $r=p^m$ for a prime $p$ and a positive integer $m \in \Z_+$.
When \eqref{dP2equation} is defined over a finite field $\F_{r}$,
the dependent variable $u_n$ will eventually take values $\pm 1$ for generic parameters and initial values $(u_0,u_1) \in \F_{r}^2$, 
and we cannot proceed to evolve it.
If we extend the domain from $\F_{r}^2$ to $(\P\F_r)^2=(\F_r\cup\{\infty\})^2$, $\P\F_r$ is not a field and we cannot define arithmetic operation in \eqref{dP2equation}. 
To determine its time evolution consistently, we have two choices:
One is to restrict the parameters and the initial values to a smaller domain so that the singularities do not appear.
The other is to extend the domain on which the equation is defined.
In this article, we will adopt the latter approach.
It is convenient to rewrite \eqref{dP2equation} as:
\begin{equation}
\left\{
\begin{array}{cl}
x_{n+1}&=\dfrac{\alpha_n}{1-x_n}+\dfrac{\beta_n}{1+x_n}-y_{n},\\
y_{n+1}&=x_n,
\end{array}
\right.
\label{dP2}
\end{equation}
where $\alpha_n:=\frac{1}{2}(z_n+a),\ \beta_n:=\frac{1}{2}(-z_n+a)$.
Then we can regard \eqref{dP2} as a mapping defined on the domain $\F_r \times \F_r$.
To resolve the indeterminacy at $x_n = \pm 1$, we apply the theory of the state of initial conditions developed by Sakai \cite{Sakai}.
First we extend the domain to $\P\F_r \times \P\F_r$, and then blow it up at four points $(x,y)=(\pm 1, \infty), (\infty, \pm 1)$ 
to obtain the space of initial conditions:
\begin{equation}
\tilde{\Omega}^{(n)}:=\mathcal{A}_{(1,\infty)}^{(n)}\cup \mathcal{A}_{(-1,\infty)}^{(n)}\cup \mathcal{A}_{(\infty,1)}^{(n)}\cup \mathcal{A}_{(\infty,-1)}^{(n)},
\label{omega}
\end{equation}
where $\mathcal{A}_{(1,\infty)}^{(n)}$ is the space obtained from the two dimensional affine space $\A^2$ by blowing up twice as
\begin{align*}
\mathcal{A}_{(1,\infty)}^{(n)}&:=\left\{ \left((x-1,y^{-1}),[\xi_1:\eta_1],[u_1:v_1]  \right)\ \Big|\ \right. \\
&\qquad \eta_1 (x-1)=\xi_1 y^{-1},
(\xi_1+\alpha_n \eta_1)v_1=\eta_1(1-x)u_1 \ \Big\} \; \subset \A^2 \times \P \times \P,
\end{align*}
where $[a:b]$ denotes a set of homogeneous coordinates for $\P^1$.
Similarly, 
\begin{align*}
\mathcal{A}_{(-1,\infty)}^{(n)}&:=\left\{ \left((x+1,y^{-1}),[\xi_2:\eta_2],[u_2:v_2]  \right)\ \Big|\ 
\right.\\
&\qquad \qquad \eta_2 (x+1)=\xi_2 y^{-1},(-\xi_2+\beta_n \eta_2)v_2=\eta_2(1+x)u_2 \ \Big\},\\
\mathcal{A}_{(\infty,1)}^{(n)}&:=\left\{ \left((x^{-1},y-1),[\xi_3:\eta_3],[u_3:v_3]  \right)\ \Big|\ 
\right. \\
& \qquad \qquad \xi_3 (y-1)=\eta_3 x^{-1}, (\eta_3+\alpha_n \xi_3)v_3=\xi_3(1-y)u_3 \ \Big\},\\
\mathcal{A}_{(\infty,-1)}^{(n)}&:=\left\{ \left((x^{-1},y+1),[\xi_4:\eta_4],[u_4:v_4]  \right)\ \Big|\ 
\right. \\
& \qquad \qquad \xi_4 (y+1)=\eta_4 x^{-1}, (-\eta_4+\beta_n \xi_4)v_3=\xi_4(1+y)u_4 \ \Big\}.
\end{align*}  
The bi-rational map \eqref{dP2} is extended to the bijection $\tilde{\phi}_n: \ \tilde{\Omega}^{(n)} \rightarrow \tilde{\Omega}^{(n+1)}$ 
which decomposes as $\tilde{\phi}_n:=\iota_n \circ \tilde{\omega}_n$. 
Here $\iota_n$ is a natural isomorphism which gives $\tilde{\Omega}^{(n)} \cong  \tilde{\Omega}^{(n+1)}$, that is,
on $\mathcal{A}_{(1,\infty)}^{(n)}$ for instance, $\iota_n$ is expressed as 
\begin{align*}
&\left((x-1,y^{-1}),[\xi :\eta ],[u :v ]  \right) \in  \mathcal{A}_{(1,\infty)}^{(n)} \\
&\rightarrow \quad
\left((x-1,y^{-1}),[\xi -\delta/2\cdot\eta:\eta ],[u :v ]  \right) \in  \mathcal{A}_{(1,\infty)}^{(n+1)}.
\end{align*}

The automorphism $\tilde{\omega}_n$ on $\tilde{\Omega}^{(n)}$ is induced from \eqref{dP2} and gives the mapping
\[
\mathcal{A}_{(1, \infty)}^{(n)} \rightarrow \mathcal{A}_{(\infty,1)}^{(n)}, \;
\mathcal{A}_{(\infty,1)}^{(n)} \rightarrow \mathcal{A}_{(-1,\infty)}^{(n)}, \;
\mathcal{A}_{(-1, \infty)}^{(n)} \rightarrow \mathcal{A}_{(\infty,-1)}^{(n)}, \;
\mathcal{A}_{(\infty,-1)}^{(n)} \rightarrow \mathcal{A}_{(1,\infty)}^{(n)}.
\]
Under the map $\mathcal{A}_{(1, \infty)}^{(n)} \rightarrow \mathcal{A}_{(\infty,1)}^{(n)}$,
\begin{align*}
x=1 \ \rightarrow \ E_2^{(\infty,1)} &\qquad u_3=\left(y-\frac{\beta_n}{2}\right)v_3, \\
E_1^{(1,\infty)} \ \rightarrow \ E_1^{(\infty,1)} &\qquad [\xi_1:-\eta_1]=[\alpha_n \xi_3+\eta_3:\xi_3], \\
E_2^{(1,\infty)} \ \rightarrow \ y'=1 &\qquad x'=\frac{u_1}{v_1}+\frac{\beta_n}{2},
\end{align*}
where $(x,y) \in \mathcal{A}_{(1, \infty)}^{(n)}$, $(x',y')\in \mathcal{A}_{(\infty,1)}^{(n)}$, $E_1^{\sp}$ and $E_2^{\sp}$ are the exceptional curves in $\mathcal{A}_{\sp}^{(n)}$ obtained by the first blowing up and the second blowing up respectively at the point p $\in \{(\pm 1, \infty),(\infty,\pm 1)  \}$. 
Similarly under the map $\mathcal{A}_{(\infty,1)}^{(n)} \rightarrow \mathcal{A}_{(-1,\infty)}^{(n)}$,
\begin{align*}
E_1^{(\infty,1)} \ \rightarrow \ E_1^{(-1,\infty)} &\qquad [\xi_3:\eta_3]=[\eta_2:(\beta_n-\alpha_n) \eta_2-\xi_2], \\
E_2^{(\infty,1)} \ \rightarrow \ E_2^{(-1,\infty)} &\qquad [u_3:v_3]=[-\beta_n u_2: \alpha_n v_2].
\end{align*}
The mapping on the other points are defined in a similar manner.
Note that $\tilde{\omega}_n$ is well-defined in the case $\alpha_n=0$ or $\beta_n=0$.
In fact, for $\alpha_n=0$, $E_2^{(1,\infty)}$ and $E_2^{(\infty,1)}$ can be identified with the lines $x=1$ and $y=1$ respectively. 
Therefore we have found that, through the construction of the space of initial conditions, the dP\II equation can be well-defined over finite fields.
However there are some unnecessary elements in the space of initial conditions when we consider a finite field, because we are working on a discrete topology and do not need continuity of the map. 
Let $\tilde{\Omega}^{(n)}$ be the space of initial conditions and $|\tilde{\Omega}^{(n)}|$ be the number of elements of it.
For the dP\II equation, we obtain $|\tilde{\Omega}^{(n)}|=(r+1)^2-4+4(r+1)-4+4(r+1)=r^2+10r+1$, since $\P\F_r$ contains $r+1$ elements.
However an exceptional curve $E_1^{\sp}$ is transferred to another exceptional curve $E_1^{\sp'}$, and $[1:0] \in E_2^{\sp}$ to 
$[1:0] \in E_2^{\sp'}$ or to a point in $E_1^{\sp'}$. Hence we can reduce the space of initial conditions $\tilde{\Omega}^{(n)}$ to the minimal space of initial conditions $\Omega^{(n)}$ which is the minimal subset of $\tilde{\Omega}^{(n)}$ including $\P\F_r\times \P\F_r$, closed under the time evolution.
By subtracting unnecessary elements we find $|\Omega^{(n)}|=(r+1)^2-4+4(r+1)-4=r^2+6r-3$.
In summary, we obtain the following proposition:
\begin{prp}
The domain of the dP\II equation over $\F_r$ can be extended to the minimal domain $\Omega^{(n)}$ on which the time evolution at time step $n$ is well defined. Moreover $|\Omega^{(n)}|=r^2+6r-3$. 
\end{prp}

\begin{figure}
\centering
\includegraphics[width=11cm,bb=-152 152 510 662]{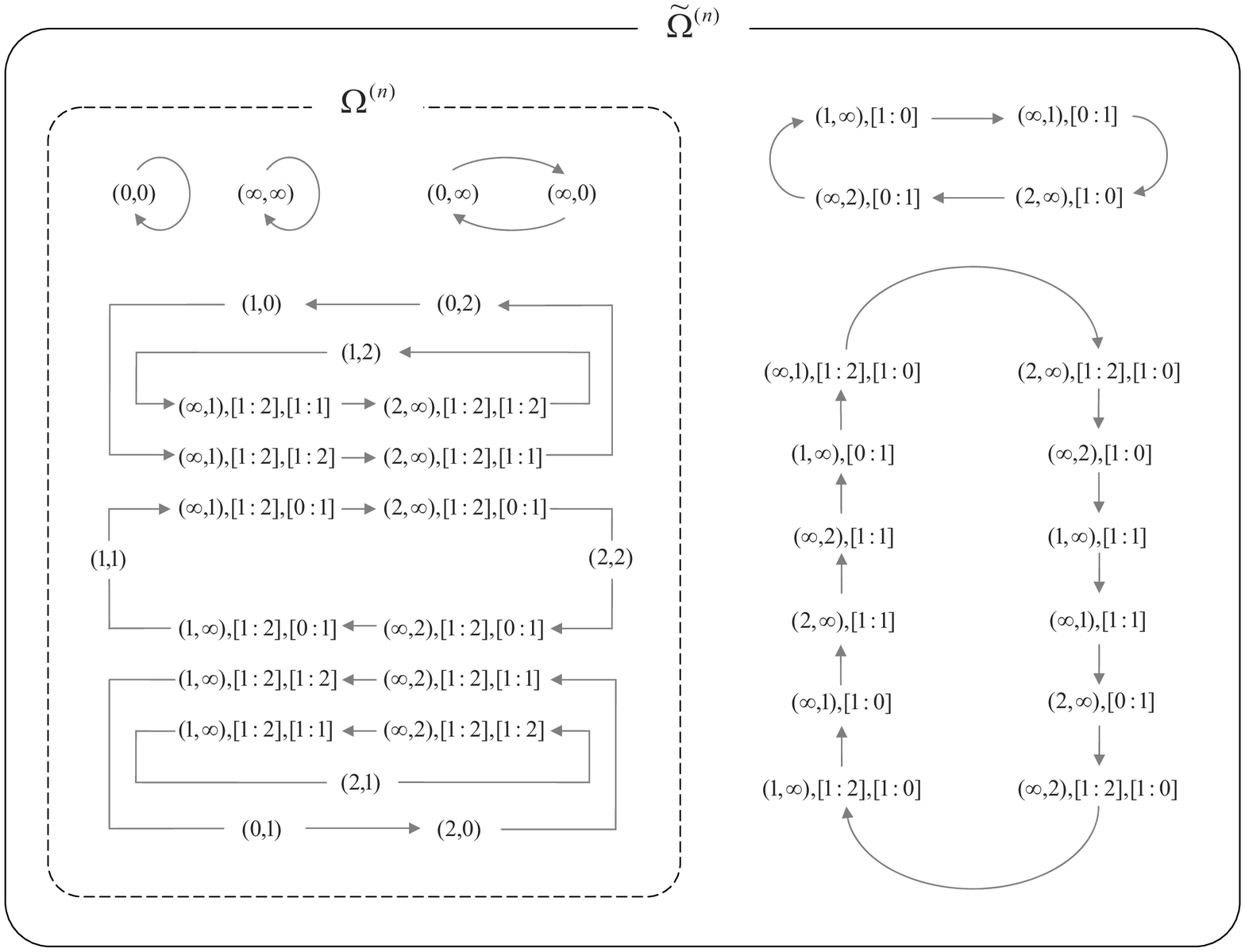}
\caption{The orbit decomposition of the space of initial conditions $\tilde{\Omega}^{(n)}$ and the reduced one $\Omega^{(n)}$ for $p=r=3$.}
\label{figure1}
\end{figure}
In figure \ref{figure1}, we show a schematic diagram of the map $\tilde{\omega}_n$ on $\tilde{\Omega}^{(n)}$, and its restriction map $\omega_n:=\tilde{\omega}_n|_{\Omega^{(n)}}$ on $\Omega^{(n)}$
with $r=3$, $\alpha_0=1$ and $\beta_0=2$.
We can also say that the figure \ref{figure1} is a diagram for the autonomous version of the equation \eqref{dP2} when $\delta=0$.
In the case of $r=3$, we have $|\tilde{\Omega}^{(n)}|=40$ and $|\Omega^{(n)}|=24$. 

The above approach is equally valid for the other discrete Painlev\'{e} equations and we can define them over finite fields by constructing isomorphisms on the spaces of initial conditions.
Thus we conclude that a discrete Painlev\'{e} equation can be well defined over a finite field by redefining the initial domain properly.
In the next section, we show another extension of the space of initial conditions: we extend it to $\Z_p\times \Z_p$\footnote{With the method in the next section, we can only deal with systems over the field $\mathbb{F}_{p^1}$. To apply it to those over the field $\mathbb{F}_{p^m}\ (m>1)$, we have to use the field extension of $\mathbb{Q}_p$.}.  
%

\section{The discrete dynamical systems over a local field and its reduction modulo a prime (review)}

\subsection{Almost good reduction}
%
Let $p$ be a prime number and for each $x \in \Q$ ($x \ne 0$) write $x=p^{v_p(x)} \dfrac{u}{v}$ where $v_p(x), u, v \in \Z$ and $u$ and $v$ are coprime integers neither of which is divisible by $p$.
The $p$-adic norm $|x|_p$ is defined as $|x|_p=p^{-v_p(x)}$. ($|0|_p=0$.)
The local field $\Q_p$ is a completion of $\Q$ with respect to the $p$-adic norm. 
It is called the field of $p$-adic numbers and its subring $\Z_p:=\{x\in \Q_p | \ |x|_p \le 1\}$ is called the ring of $p$-adic integers \cite{Murty}. 
The $p$-adic norm satisfies a non-Archimedean (ultrametric) triangle inequality 
$|x+y|_p \le \max[|x|_p,|y|_p ]$.
Let $\mathfrak{p}=p\Z_p=\left\{x \in \Z_p |\ v_p(x) \ge 1 \right\}$ be the maximal ideal of $\Z_p$.
We define the reduction of $x$ modulo $\mathfrak{p}$ as
\[
\tilde{x}: \Z_p \ni x \mapsto \tilde{x}:=x\mod \mathfrak{p} \in \Z_p/\mathfrak{p} \cong \F_p.
\]
Note that the reduction is a ring homomorphism.
The reduction is generalised to $\Q_p^{\times}$:
\[
\Q_p^{\times}\ni x=p^k u\ (u\in\Z_p^{\times})\mapsto
\left\{
\begin{array}{cl}
\tilde{x} & (k\ge0)\\
\infty & (k<0)
\end{array}
\right. \in \P\F_p,
\]
which is no longer homomorphic. Note that for $x\in\Q_p^{\times}$, $k\ge 0$ is equivalent to $x\in\Z_p^{\times}$.
Let us consider a rational system $\phi$ consisting of two rational functions of $x$ and $y$:
\[
\phi(x,y)=(\phi_1(x,y),\phi_2(x,y))\in(\mathbb{Z}_p(x,y))^2: \Q_p^2 \to \Q_p^2,
\]
its reduction
\[
\tilde{\phi}(x,y)=(\tilde{\phi}_1(x,y),\tilde{\phi}_2(x,y))\in(\mathbb{F}_p(x,y))^2
\]
is defined as the system whose coefficients are all reduced.
The rational system is said to have a \textit{good reduction} (modulo $\mathfrak{p}$ on the domain $\mathcal{D}$) if we have $\widetilde{\phi(x,y)}=\tilde{\phi}(\tilde{x},\tilde{y})$ for any $(x,y) \in \mathcal{D}$ \cite{Silverman}.
We have defined a generalised notion in our previous letter and have explained its usefulness;
\begin{dfn}[\cite{KMTT}]
A (non-autonomous) rational system $\phi_n$: $\Q_p^2 \to \Q_p^2$ $(n \in \Z)$ has an almost good reduction modulo $\mathfrak{p}$ on the domain $\mathcal{D}\subseteq \Z_p^2$, if there
exists a positive integer $m_{\mbox{\rm \scriptsize p};n}$ for any $\mbox{\rm p}=(x,y) \in \mathcal{D}$ and time step $n$ such that
\begin{equation}
\widetilde{\phi_n^{m_{\mbox{\rm \tiny p};n}}(x,y)}=\widetilde{\phi_n^{m_{\mbox{\rm \tiny p};n}}}(\tilde{x},\tilde{y}),
\label{AGR}
\end{equation}
where $\phi_n^m :=\phi_{n+m-1} \circ \phi_{n+m-2} \circ \cdots \circ \phi_n$.
\end{dfn} 
Let us first review some of the findings in \cite{KMTT} in order to see the significance of the notion of \textit{almost good reduction}. Let us consider the mapping $\Psi_\gamma$:
\begin{equation}
\left\{
\begin{array}{cl}
x_{n+1}&=\dfrac{ax_n+1}{x_n^\gamma y_n},\\
y_{n+1}&=x_n,
\end{array}
\right.
\label{discretemap}
\end{equation} 
where $a \in \{1,2,\cdots, p-1\}$  and $\gamma \in \Z_{\ge 0}$ are parameters. 
The map \eqref{discretemap} is known to be integrable if and only if $\gamma=0,1,2$.
Note that when $\gamma=0,1,2$, \eqref{discretemap} is an autonomous version of the $q$-discrete Painlev\'{e} I equation and therefore is integrable.
We also note that when $\gamma=2$, \eqref{discretemap} belongs to the QRT family and is integrable in the sense that it has a conserved quantity.

Let $\mathcal{D}$ be the domain $\{(x,y) \in \Z_p^2 \ |x \ne 0, y \ne 0\}$, then clearly 
\[
\widetilde{\Psi_2(x_n,y_n)}=\widetilde{\Psi}_2(\tilde{x}_n,\tilde{y}_n) \qquad \mbox{for $\tilde{x}_n \ne 0, \ \tilde{y}_n \ne 0$}.
\] 
For $(x_n,y_n)\in\mathcal{D}$ with $\tilde{x}_n=0$ and $\tilde{y}_n \ne 0$, we find that
$\widetilde{\Psi_2^k}(\tilde{x}_n=0,\tilde{y}_n)$ is not defined for $k=1,2$,
however it is defined if $k=3$ and we have
\[
\widetilde{\Psi_2^3(x_n,y_n)}=\widetilde{\Psi_2^3}(\tilde{x}_n=0,\tilde{y}_n)=\left(\dfrac{1}{a^2\tilde{y}},0\right) .
\]
Finally for $\tilde{x}_n=\tilde{y}_n = 0$, we find that
$\widetilde{\Psi_2^k}(\tilde{x}_n,\tilde{y}_n)$ is not defined for $k=1,2,..,7$,
however
\[
\widetilde{\Psi_2^8(x_n,y_n)}=\widetilde{\Psi_2^8}(\tilde{x}_n=0,\tilde{y}_n=0)=\left(0,0\right) .
\]
Hence the map $\Psi_2$ has almost good reduction modulo $\mathfrak{p}$ on $\mathcal{D}$.
Note that, in the case $\gamma=2$ and $a=0$,
if we take 
\[
f_{2k}:=x_{2k}x_{2k-1},\ f_{2k-1}:=(x_{2k-1}x_{2k-2})^{-1}
\]
\eqref{discretemap} turns into the trivial linear mapping $f_{n+1}=f_n$ which has apparently good reduction modulo $\mathfrak{p}$.
In a similar manner, we find that $\Psi_\gamma$ ($\gamma=0,1$) also has almost good reduction modulo $\mathfrak{p}$ on $\mathcal{D}$. 
On the other hand, for $\gamma \ge 3$ and $\tilde{x}_n=0$, we easily find that
\[
{}^\forall k \in \Z_{\ge 0}, \;\; \widetilde{\Psi_\gamma^{k}(x_n,y_n)} \ne \widetilde{\Psi_\gamma^{k}}(\tilde{x}_n=0,\tilde{y}_n),
\]
since the order of $p$ diverges as we iterate the mapping.
Thus we have proved the following proposition:
\begin{prp}
The rational mapping \eqref{discretemap} has an almost good reduction modulo $\mathfrak{p}$ only for $\gamma=0,1,2$.
\label{PropQRT}
\end{prp}
Note that having an almost good reduction is equivalent to the integrability of the equation in these examples.

\subsection{The dP\II equation modulo a prime and its special solutions}

Now let us examine the dP\II \eqref{dP2} over $\Q_p$. 
We suppose that $p \ge 3$, and redefine the coefficients $\alpha_n$ and $\beta_n$ so that
they are periodic with period $p$:
\begin{align*}
\alpha_{i+mp}&:=\frac{(i\delta+z_0+a+n_\alpha p)}{2},\ \beta_{i+mp}:=\frac{(-i\delta-z_0+a+n_\beta p)}
{2},\\
(m\in\Z&,\  i\in \{0,1,2,\cdots,p-1\}),
\end{align*}
where the integer $n_\alpha$ ($n_\beta$) is chosen such that $0 \in \{\alpha_i\}_{i=0}^{p-1}$ $(0 \in \{\beta_i\}_{i=0}^{p-1})$.
As a result, we have $\tilde{\alpha}_{n}=\widetilde{\frac{n \delta +z_0+a}{2}}$, $\tilde{\beta}_{n}=\widetilde{\frac{-n \delta -z_0+a}{2}}$ and $|\alpha_n|_p,\ |\beta_n|_p\in\{0,1\}$ for any integer $n$.

\begin{prp}
Under the above assumptions, the dP\II equation has an almost good reduction modulo $\mathfrak{p}$ on $\mathcal{D}:=\{ (x,y) \in \Z_p^2\ |x \ne \pm 1\}$.
\label{PropdP2}
\end{prp}
\textbf{proof}\;\; 
We put $(x_{n+1},y_{n+1})=\phi_n(x_n,y_n)=\left( \phi_n^{(x)}(x_n,y_n),\phi_n^{(y)} (x_n,y_n) \right)$. 

When $\tilde{x}_n \ne \pm 1$, we have
\[
\tilde{x}_{n+1}=\dfrac{\tilde{\alpha_n}}{1-\tilde{x}_n}+\dfrac{\tilde{\beta_n}}{1+\tilde{x}_n}-\tilde{y}_{n},
\quad \tilde{y}_{n+1}=\tilde{x}_n.
\]
Hence $\widetilde{\phi_n(x_n,y_n)}=\tilde{\phi}_n(\tilde{x}_n,\tilde{y}_n)$.

When $\tilde{x}_n=1$, we can write $x_n=1+p^k e$ $(k \in \Z_+,\ |e|_p=1)$. 
We have to consider four cases\footnote{Precisely speaking, there are some special cases for $p=3,5$ where we have to consider the fact $\alpha_{n+p}=\alpha_{n}$ or $\beta_{n+p}=\beta_{n}$ during the iteration process. We can prove by straightforward calculations that similar results hold for these exceptional cases.}:\\
\noindent
(i) For $\alpha_n= 0 $,
\[
\tilde{x}_{n+1}=\tilde{\phi}_n^{(x)}(\tilde{x}_n,\tilde{y}_n)=\widetilde{\left(\frac{\beta_n}{2}\right)}-\tilde{y}_n.
\]
Hence we have $\widetilde{\phi_n(x_n,y_n)}=\tilde{\phi}_n(\tilde{x}_n,\tilde{y}_n)$.\\
(ii) In the case $\alpha_n \neq 0$ and $\beta_{n+2} \neq 0$, 
\begin{align*}
x_{n+1}&=-\dfrac{(\alpha_n-\beta_n)(1+ep^k)+a}{ep^k(2+ep^k)}-y_n=-\dfrac{2\alpha_n+(\alpha_n-\beta_n)ep^k}{ep^k(2+ep^k)}-y_n,\\
x_{n+2}&=-\frac{\alpha_n^2+\mbox{polynomial of $O(p)$}}{\alpha_n^2+\mbox{polynomial of $O(p)$}},\\
x_{n+3}&=\dfrac{\{2\alpha_{n}y_n+2\delta \beta_{n+1}+(2-\delta)a \}\alpha_n^3 +\mbox{polynomial of $O(p)$}}{2\beta_{n+2}
\alpha_n^3 + \mbox{polynomial of $O(p)$} },
\end{align*}
Thus we have
\[
\tilde{x}_{n+3}=\frac{2\tilde{\alpha}_{n}\tilde{y}_n+2\delta \tilde{\beta}_{n+1}+(2-\delta)a}{2 \tilde{\beta}_{n+2}},
\quad \tilde{y}_{n+3}=-1,
\]
and $\widetilde{\phi_n^3(x_n,y_n)}=\widetilde{\phi_n^3}(\tilde{x}_n,\tilde{y}_n)$.\\
(iii) In the case $\alpha_n \neq 0$, $\beta_{n+2}= 0$ and $a \ne -\delta$, we have to calculate up to $x_{n+5}$.
After a lengthy calculation we find
\[
\tilde{x}_{n+4}=\widetilde{\phi_n^{5}}^{(y)}(1,\tilde{y}_n)=1,\;\mbox{and}\;
\tilde{x}_{n+5}=\widetilde{\phi_n^{5}}^{(x)}(1,\tilde{y}_n)=-\frac{a\delta-(a-\delta)\tilde{y}_n}{a+\delta},
\]  
and we obtain $\widetilde{\phi_n^5(x_n,y_n)}=\widetilde{\phi_n^5}(\tilde{x}_n,\tilde{y}_n)$.\\
(iv) Finally, in the case $\alpha_n \neq 0$, $\beta_{n+2}= 0$ and $a = -\delta$ we have to calculate up to $x_{n+7}$.
The result is
\[
\tilde{x}_{n+6}=\widetilde{\phi_n^{7}}^{(y)}(1,\tilde{y}_n)=-1,\;\mbox{and}\;
\tilde{x}_{n+7}=\widetilde{\phi_n^{7}}^{(x)}(1,\tilde{y}_n)=\frac{1+2\tilde{y}_n}{2},
\]
and we obtain $\widetilde{\phi_n^7(x_n,y_n)}=\widetilde{\phi_n^7}(\tilde{x}_n,\tilde{y}_n)$.
Hence we have proved that the dP\II equation has almost good reduction modulo $\mathfrak{p}$ at $\tilde{x}_n=1$.
We can proceed in the case $\tilde{x}_n=-1$ in an exactly similar manner. $\Box$

From this proposition, the evolution of the dP\II equation \eqref{dP2equation} over $\P\F_p$ can be constructed from the initial values $u_{n-1}$ and $u_n$.

Now we consider special solutions to \eqref{dP2equation} over $\P\F_p$.
For the dP\II equation over $\C$, rational function solutions have already been obtained \cite{Kajiwara}.
Let $N$ be a positive integer and $\lambda \ne 0$  be a constant. Suppose that 
$
a=-\frac{2(N+1)}{\lambda}$, $\quad \delta=z_0=\frac{2}{\lambda}$,
\[
L_k^{(\nu)}(\lambda):=\left\{ 
\begin{array}{cl}
\displaystyle
\sum_{r=0}^k(-1)^r
\begin{pmatrix}
k+\nu\\
k-r
\end{pmatrix} 
\dfrac{\lambda^r}{r!}&\quad(k \in \Z_{\ge 0}),\\
0 &\quad (k \in \Z_{<0}),
\end{array}
\right.
\]
and
\begin{equation}
\tau_N^n:=\det
\begin{vmatrix}
L_{N+1-2i+j}^{(n)}(\lambda)
\end{vmatrix}_{1\le i,j\le N}.
\label{Ltau}
\end{equation}
Then a rational function solution of the dP\II equation is given by
\begin{equation}
u_n=\frac{\tau_{N+1}^{n+1}\tau_{N}^{n-1}}{\tau_{N+1}^n\tau_N^n}-1.
\label{rationaldP2}
\end{equation}
If we deal with the terms in \eqref{Ltau} and \eqref{rationaldP2} by arithmetic operations over $\F_p$, 
we encounter terms such as $\frac{1}{p}$ or $\frac{p}{p}$ and \eqref{rationaldP2} is not well-defined.
However, from proposition \ref{PropdP2}, we find that \eqref{rationaldP2} gives a solution to the dP\II equation over $\P\F_q$ by
the reduction from $\Q \ (\subset \Q_p)$, as long as the solution avoids the points $(\tilde{\alpha}_n=0,\ u_n=1)$ and $(\tilde{\beta_n}=0,\ u_n=-1)$, which is equivalent to the solution satisfying
\begin{equation}
\tau_{N+1}^{-N-1} \tau_N^{-N-3}\not\equiv 0,\ \frac{\tau_{N+1}^{N+1}\tau_N^{N-1}}{\tau_{N+1}^N\tau_N^N}\not\equiv 2, \label{taucond}
\end{equation}
where the superscripts are considered modulo $p$. Note that $\tau_N^n\equiv \tau_N^{n+p}$ for all integers $N$ and $n$.
In the table below, we give several \textit{rational solutions to the dP\II equation} with $N=3$ and $\lambda=1$ over $\P\F_q$ for $q=3,5,7$ and $11$. We see that the period of the solution is $p$.
\begin{scriptsize}
\[
\begin{array}{|c|c|c|l|}
\hline
& & & \\[-2mm]
\raise3mm\hbox{$p$} & \raise3mm\hbox{$\tau_{N+1}^{-N-1}\tau_N^{-N-3}$} & \raise3mm\hbox{$\frac{\tau_{N+1}^{N+1}\tau_N^{N-1}}{\tau_{N+1}^N\tau_N^N}$}
&\enskip
\raise3mm\hbox{$\tilde{u}_1,\tilde{u}_2,\tilde{u}_3,\tilde{u}_4,\tilde{u}_5,\tilde{u}_6,\tilde{u}_7,\tilde{u}_8,\tilde{u}_9,\tilde{u}_{10},\ldots$} \\ \hline & & & \\[-2mm]
3 & \infty & \infty &\
\raise1mm\hbox{$\underbrace{1,2,\infty}_{\mbox{period
$3$}},1,2,\infty,1,2,\infty,1,2,\infty,1,2,\infty,\ldots$}
\\[5mm] \hline
& & & \\[-2mm]
5 & \infty & 4 &\
\raise1mm\hbox{$\underbrace{4,2,3,1,\infty}_{\mbox{period
$5$}},4,2,3,1,\infty,4,2,3,1,\infty,4,\ldots$} \\[5mm] \hline & & & \\[-2mm]
7 & \infty & 0 &\
\raise1mm\hbox{$\underbrace{1,\infty,6,5,1,\infty,6}_{\mbox{period
$7$}},1,\infty,6,5,1,\infty,6,1,\infty,\ldots$} \\[5mm] \hline & & & \\[-2mm]
11 & 0 & 7 &\
\raise1mm\hbox{$\underbrace{\infty,1,6,1,\infty,10,\infty,1,0,2,10}_{\mbox{period
$11$}},\infty,1,6,1,\ldots$} \\[5mm] \hline \end{array} \]
\end{scriptsize}
We see from the case of $p=11$ that we may have an appropriate solution even if the condition \eqref{taucond} is not satisfied, although this is not always true.
The dP\II equation has linearized solutions also for $\delta=2a$ \cite{Tamizhmani}. 
With our new method, we can obtain the corresponding solutions without difficulty.

\section{The $p$-adic analogue of the singularity confinement test}

The above approach is closely related to the singularity confinement method which is an effective test to judge the integrability of the given equations \cite{Grammaticosetal}.
In the proof of the Proposition \ref{PropdP2}, we have taken $x_n=1+e p^k$ and have shown that the limit \[\lim_{|e p^k|_p \to 0}(x_{n+m}, x_{n+m+1})\]
is well defined for some positive integer $m$.
Here $ep^k\ (k>0,\ |e|_p=1)$ is an alternative in $\Q_p$ for the infinitesimal parameter $\epsilon$ in the singularity confinement test in $\C$. Note that $p^k\ (k>0)$ is a `small' number in terms of the $p$-adic metric.
From this observation and propositions \ref{PropQRT}, \ref{PropdP2}, we postulate that having almost good reduction in arithmetic mappings is similar to passing the singularity confinement test.
%
%

\section{The $q$-discrete Painlev\'{e} equations over a finite field}
%
%
In this section we study the $q$-discrete analogues of the Painlev\'{e} equations ($q$P\I and $q$P\II equations).

\subsection{The $q$P\I equation}
One of the forms of the $q$P\I equations is as follows:
\begin{equation}
x_{n+1}x_{n-1}=\frac{aq^nx_n+b}{x_n^2}, \label{qp1eq}
\end{equation}
where $a$ and $b$ are parameters.
We rewrite \eqref{qp1eq} for our convenience as
\begin{equation}
\Phi_n: \left\{
\begin{array}{cl}
x_{n+1}&=\dfrac{aq^n x_n+b}{x_n^2 y_n},\\
y_{n+1}&=x_n.
\end{array}
\right.
\label{qP1}
\end{equation} 
Similarly to the dP\II equation, we can prove the following proposition:
\begin{prp}
Suppose that $a, b, q$ are integers not divisible by $p$, then the mapping \eqref{qP1} has an almost good reduction   
modulo $\mathfrak{p}$ on the domain 
$\mathcal{D}:=\{(x,y)\in \Z_p^2\ |x \ne 0, y\ne 0\}$.
\label{PropqP1}
\end{prp}
\textbf{proof}\;\;
Let $(x_{n+1},y_{n+1})=\Phi_n(x_n,y_n)$.
Just like we have done before, we have only to examine the cases $\tilde{x}_n=0$ and $\tilde{y}_n=0$.
We use the abbreviation $\tilde{q}=q, \tilde{a}=a, \tilde{b}=b$ for simplicity. By direct computation we obtain;

(i) If $\tilde{x}_n=0$ and $\tilde{y}_n\ne 0$, then
\[
\widetilde{\Phi_n^3(x_n,y_n)} = \widetilde{\Phi_n^3}(0,\tilde{y}_n)=\left(\frac{b^2}{a^2 q^2 \tilde{y}_n},0\right).
\]

(ii) If $\tilde{y}_n=0$ and $\tilde{x}_n\ne 0$, then
\[
\widetilde{\Phi_n^5(x_n,y_n)} = \widetilde{\Phi_n^5}(\tilde{x}_n,0)=\left(0, \frac{a^2 q^4}{b \tilde{x}_n}\right).
\]

(iii) If $\tilde{x}_n=0$ and $\tilde{y}_n= 0$, then
\[
\widetilde{\Phi_n^8(x_n,y_n)} = \widetilde{\Phi_n^8}(0,0)=\left( 0, 0\right).
\]

\subsection{The $q$P\II equation}
The $q$P\II equation is the following $q$-discrete equation:
\begin{equation}
(z(q\tau)z(\tau)+1)(z(\tau)z(q^{-1}\tau)+1)=\frac{a \tau^2 z(\tau)}{\tau-z(\tau)},
\label{qP2eq}
\end{equation}
where $a$ and $q$ are parameters \cite{Kajiwaraetal}.
It is also convenient to rewrite \eqref{qP2eq} as
\begin{equation}
\Phi_n: \left\{
\begin{array}{cl}
x_{n+1}&=\dfrac{a(q^n\tau_0)^2x_n-(q^n\tau_0-x_n)(1+x_ny_n)}{x_n(q^n\tau_0-x_n)(x_ny_n+1)},\\
y_{n+1}&=x_n,
\end{array}
\right.
\label{qP2}
\end{equation}
where $\tau=q^n\tau_0$.
Just like the qP\I equation, we can prove that qP\II has an almost good reduction:
\begin{prp}
Suppose that $a, q, \tau_0$ are integers not divisible by $p$, then the mapping \eqref{qP2} has an almost good reduction   
modulo $\mathfrak{p}$ on the domain 
$\mathcal{D}:=\{(x,y)\in \Z_p^2\ |x \ne 0, x \ne q^n\tau_0\ (n\in\Z), xy+1 \ne 0\}$.
\label{PropqP2}
\end{prp}
\textbf{proof}\;\;
Let $(x_{n+1},y_{n+1})=\Phi_n(x_n,y_n)$.
Just like we have done before, we have only to examine the cases $\tilde{x}_n=0, \widetilde{q^n\tau_0}$ 
and $-\tilde{y}_n^{-1}$.
 We obtain;\\
(i) If $\tilde{x}_n=0$ and $ -1+q^2-aq^2\tau^2+q^3\tau^2-q^2\tau \tilde{y}_n \ne 0$, 
\[
\widetilde{\Phi_n^3(x_n,y_n)} = \widetilde{\Phi_n^3}(\tilde{x}_n=0,\tilde{y}_n)=\left( \frac{ 1-q^2+aq^2\tau^2-q^3\tau^2-aq^4\tau^2+q^2\tau \tilde{y}_n}{q^2\tau( -1+q^2-aq^2\tau^2+q^3\tau^2-q^2\tau \tilde{y}_n ) }   , q^2\tau  \right).
\]
(ii) If $\tilde{x}_n=0$ and $ -1+q^2-aq^2\tau^2+q^3\tau^2-q^2\tau \tilde{y}_n = 0$, 
\[
\widetilde{\Phi_n^5(x_n,y_n)} = \widetilde{\Phi_n^5}(\tilde{x}_n=0,\tilde{y}_n) =\left( \frac{1-q^2+q^7\tau^2-aq^8\tau^2}{q^4\tau}, 0    \right).
\]
(iii) If $\tilde{x}_n=\tau$ and $1+\tau \tilde{y}_n\ne 0$,
\begin{align*}
&\widetilde{\Phi_n^3(x_n,y_n)} = \widetilde{\Phi_n^3}(\tilde{x}_n=\tau,\tilde{y}_n)\\
&\quad =\left(\frac{ 1-q^2+(a+q-aq^2)q^2\tau^2+(1-q^2)\tau\tilde{y}+(1-aq)q^3\tau^3\tilde{y}  }{q^2\tau(1+\tau\tilde{y}_n)}, 0 \right).
\end{align*}
(iv) If $\tilde{x}_n=\tau$ and $1+\tau \tilde{y_n}= 0$,
\[
\widetilde{\Phi_n^7(x_n,y_n)} = \widetilde{\Phi_n^7}(\tilde{x}_n=\tau,\tilde{y}_n)=\left(\frac{1}{aq^{12}\tau^3}, - aq^{12}\tau^3  \right).
\]
(v) If $\tilde{x_n}\tilde{y}_n+1=0$,
\[
\widetilde{\Phi_n^7(x_n,y_n)} = \widetilde{\Phi_n^7}(\tilde{x}_n=-\tilde{y}_n^{-1}, \tilde{y}_n)=\left(-\frac{1}{aq^{12}\tau^4\tilde{y}_n}, aq^{12}\tau^4\tilde{y}_n  \right).
\]
Thus we complete the proof. $\Box$

From this proposition we can define the time evolution of the $q$P\II equation explicitly just like the dP\II equation in the previous section.

Next we consider special solutions for qP\II equation \eqref{qP2eq} over $\P\F_p$. 
In \cite{HKW} it has been proved that \eqref{qP2eq} over $\C$ with $a=q^{2N+1}$ $(N \in \Z)$ is solved by
the functions given by
\begin{eqnarray}
z^{(N)} (\tau) &= 
\begin{cases}
\displaystyle \frac{g^{(N)} (\tau) g^{(N+1)} (q \tau)}{q^N g^{(N)} (q \tau) g^{(N+1)} (\tau)}
 & (N \ge 0), \\
\displaystyle \frac{g^{(N)} (\tau) g^{(N+1)} (q \tau)}{q^{N+1} g^{(N)} (q \tau) g^{(N+1)} (\tau)} & (N<0),
\end{cases} \label{eq:gtoz} \\
g^{(N)} (\tau) &= 
\begin{cases}
\begin{vmatrix}
w(q^{-i+2j-1}\tau)
\end{vmatrix}_{1\le i,j\le N} & (N>0), \\
\ 1 & (N=0), \\
\begin{vmatrix}
w(q^{i-2j} \tau)
\end{vmatrix}_{1\le i,j\le -N} & (N<0),
\end{cases} \label{eq:det_sol_g}
\end{eqnarray}
where $w(\tau)$ is a solution of the $q$-discrete Airy equation:
\begin{equation}
w(q\tau)-\tau w(\tau)+w(q^{-1}\tau)=0. 
\label{dAiryeq}
\end{equation}
As in the case of the dP\II equation, we can obtain the corresponding solutions 
to \eqref{eq:gtoz} over $\P\F_p$ by reduction modulo $\mathfrak{p}$ according to the proposition \ref{PropqP2}.
For that purpose, we have only to solve \eqref{dAiryeq} over $\Q_p$.
By elementary computation we obtain:
\begin{equation}
w(q^{n+1}\tau_0)=c_1P_{n}(\tau_0;q)+c_0P_{n-1}(q\tau_0;q),
\label{Airy:sol}
\end{equation}
where $c_0,\ c_1$ are arbitrary constants and $P_n(x;q)$ is defined by the tridiagonal determinant: 
\[
P_n(x;q):=
\left|
\begin{array}{ccccc}
qx&-1&&&\\
-1&q^2x&-1&&\hugesymbol{0}\\
&\ddots&\ddots&\ddots& \\
&&-1&q^{n-1}x&-1\\
\hugesymbol{0}&&&-1&q^{n}x
\end{array}
\right|.
\]
The function $P_n(x;q)$ is the polynomial of $n$th order in $x$,
\[
P_n(x;q)=\sum_{k=0}^{[n/2]}(-1)^k a_{n;k}(q)x^{n-2k},
\]
where $a_{n;k}(q)$ are polynomials in $q$.
If we let $i \ll j $ denotes $i<j-1$, and 
$
c(j_1,j_2,...,j_k):=\sum_{r=1}^k (2j_r+1),
$
then, we have
\[
a_{n;k}=\sum_{1\le j_1 \ll j_2 \ll \cdots \ll j_k \le n-1} q^{n(n+1)/2 -c(j_1,j_2,...,j_k)}.
\] 
Therefore the solution of $q$P\II equation over $\P\F_p$ is obtained by reduction modulo $\mathfrak{p}$ from \eqref{eq:gtoz}, \eqref{eq:det_sol_g}
and \eqref{Airy:sol} over $\Q$ or $\Q_p$.

We can also prove that a type of $q$-discrete Painlev\'{e} III, IV and V equations have an almost good reduction on some appropriate domains.
These facts justifies that almost good reduction can be a criterion for integrability of systems over finite fields.
Proofs are fairly straightforward and will be reported  in the future article \cite{Kanki}.

\section{The discrete KdV equation}
In section \ref{sec2}, we have successfully determined the time evolution of the dP\II equation through the construction of the space of initial conditions of the dP\II equation by blowing-up twice at each of the singular points so that the mapping becomes bijective.
However, for a general nonlinear equation, explicit construction of the space of initial conditions over a finite field is not so straightforward (for example see \cite{Takenawa}) and it will not help us to obtain the explicit solutions.

In this section we study the soliton equations evolving as a two-dimensional lattice over finite fields by following the discussions made in \cite{KMT}.
For example, let us consider the discrete KdV equation
\begin{equation}
\frac{1}{x_{n+1}^{t+1}}-\frac{1}{x_n^t}+\frac{\delta}{1+\delta}\left(x_n^{t+1}-x_{n+1}^t \right)=0,
\label{dKdV1}
\end{equation}
over a finite field $\F_r$ where $r=p^m$, $p$ is a prime number and $m\in\Z_{+}$. Here $n,t \in \Z$ and $\delta$ is a parameter.
If we put
\[
\frac{1}{y_n^t}:=(1+\delta)\frac{1}{x_n^{t+1}}-\delta x_n^t
\]
we obtain equivalent coupled equations
\begin{equation}
\left\{
\begin{array}{cl}
x_n^{t+1}&=\dfrac{(1+\delta)y_n^t}{1+\delta x_n^ty_n^t},\vspace{2mm} \\
y_{n+1}^{t}&=\dfrac{(1+\delta x_n^ty_n^t)x_n^t}{1+\delta}.
\end{array}
\right.
\label{dKdV2}
\end{equation}
Clearly \eqref{dKdV2} does not determine the time evolution when $1+\delta x_n^t y_n^t\equiv 0$.
Over a field of characteristic 0 such as $\C$, the time evolution of $(x_n^t,y_n^t)$ will not hit this exceptional line 
for generic initial conditions, but on the contrary, the evolution comes to this exceptional line in many cases over a finite field as a division by $0$ appears.
The mapping, $(x_n^t,y_n^t) \mapsto (x_n^{t+1}, y_{n+1}^t)$, is lifted to an automorphism of the surface $\tilde{X}$,
where $\tilde{X}$ is obtained from $\P^1 \times \P^1$ by blowing up twice at $(0,\infty)$ and $(\infty, 0)$ respectively:
\begin{align*}
\tilde{X}&=\A_{(0,\infty)} \cup \A_{(\infty,0)},\\
\A_{(0,\infty)}&:=\left\{ \left(\left(x, y^{-1}\right), [\xi:\eta], [u:v]\right)  \Big| \ x \eta=y^{-1} \xi,\  \eta u = y^{-1} (\eta+\delta \xi) v \right\}\subset \A^2 \times \P^1\times\P^1, \\
\A_{(\infty,0)}&:=\left\{ \left(\left(x^{-1}, y\right), [\xi:\eta], [w:z]\right)  \Big|\ x^{-1}\eta=y\xi,\ (\eta + \delta \xi) u = y \eta v\right\}\subset \A^2 \times \P^1\times\P^1.
\end{align*}
To define the time evolution of the system with $N$ lattice points from \eqref{dKdV2}, however, we have to consider the mapping 
\[
(y_1^t;x_1^t,x_2^t,...,x_N^t) \longmapsto  (x_1^{t+1},x_2^{t+1},...,x_N^{t+1};y_{N+1}^t).
\]
Since there seems no reasonable decomposition of $\tilde{X}$ into a direct product of two independent spaces, successive use of \eqref{dKdV2} becomes impossible.
Note that if we blow down $\tilde{X}$ to $\P^1 \times \P^1$, the information of the initial values is lost in general.
If we intend to construct an automorphism of a space of initial conditions, it will be inevitable to start from $\P^{N+1}$ and blow-up to some huge manifold, which is beyond the scope of the present paper.
This difficulty seems to be one of the reasons why the singularity confinement method has not been used for construction of integrable partial difference equations or judgement for their integrability, though some attempts have been proposed in the bilinear form \cite{RGS}.
There should be so many exceptional hyperplanes in the space of initial conditions (if it does exist), and it is practically impossible to check all the ``singular'' patterns in the na\"{i}ve extension of the singularity confinement test.
On the other hand, when we fix the initial condition for a partial difference equation, the number of singular patterns is restricted in general and we have only to enlarge the domain so that the mapping becomes well-defined.
This is the strategy that we will adopt in this section.

First we explain how the indeterminate values appear through the time evolution by examining the discrete KdV equation \eqref{dKdV2} over $\F_7:=\{0,1,2,3,4,5,6\}$.
If we take $\delta=1$, \eqref{dKdV2} turns into
\[
\left\{
\begin{array}{cl}
x_n^{t+1}&=\dfrac{2y_n^t}{1+x_n^ty_n^t},\vspace{2mm} \\
y_{n+1}^t&=\dfrac{(1+x_n^ty_n^t)x_n^t}{2}.
\end{array}
\right.
\]
Suppose that $x_1^0=6,\ x_2^0=5,\ y_1^0=2,\ y_1^1=2$, then 
we have 
\[
x_1^{1}=\frac{4}{13} \equiv 3,\quad y_2^{0}=\frac{78}{2} \equiv 4 \ \mod 7.
\]
With further calculation we have
\[
x_1^2=\frac{4}{7} \equiv \frac{4}{0},\quad y_2^1=\frac{21}{2} \equiv 0,\quad x_2^1=\frac{8}{21} \equiv \frac{1}{0}.
\]
Since $\dfrac{4}{0}$ and $\dfrac{1}{0}$ are not defined over $\F_7$, we now extend $\F_7$ to $\P\F_7$ and 
take $\dfrac{j}{0}\equiv \infty$ for $j\in\{1,2,3,4,5,6\}$.
However, at the next time step, we have
\[
x_2^2=\frac{2 \cdot 0}{1+ \infty \cdot 0},\qquad y_3^1=\frac{(1+ \infty \cdot 0)\cdot \infty}{2}
\]
and reach a deadlock.

\begin{figure}
\centering
\includegraphics[width=10cm,bb=80 570 440 740]{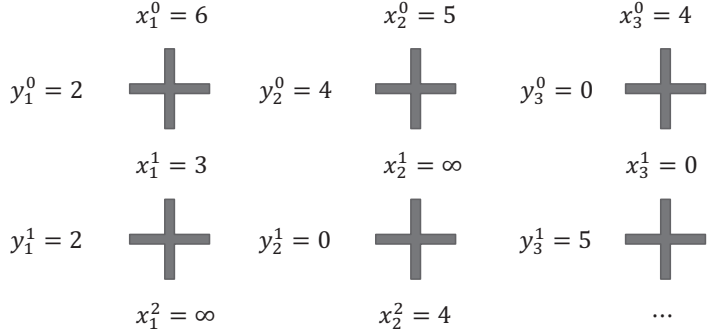}
\caption{An example of the time evolution of the coupled discrete KdV equation \eqref{dKdV2} over $\P\F_7$ where $\delta=1$.}
\label{figurekdv1}
\end{figure}

Therefore we try the following two procedures:
[I] we keep $\delta$ as a parameter for the same initial condition, and obtain as a system over $\F_7(\delta)$,
\begin{align*}
x_1^{1}&=\frac{2(1+\delta)}{1+5\delta},\quad y_2^{0}=\frac{6(1+5\delta)}{1+\delta},\\
x_2^1&=\frac{6(1+\delta)(1+5\delta)}{1+3\delta+3\delta^2},
\quad y_2^1=\frac{2(1+2\delta+4\delta^2)}{(1+5\delta)^2},\quad
x_1^2=\frac{2(1+\delta)(1+5\delta)}{1+2\delta+4\delta^2},\\
x_2^2&=\frac{4(1+\delta)(2+\delta)(3+2\delta)}{(1+5\delta)(5+5\delta+2\delta^2)},\quad y_3^1=\frac{2(5+5\delta+2\delta^2)}{(2+\delta)^2}.
\end{align*}
[II] Then we put $\delta=1$ to have a system over $\P\F_7$ as
\begin{align*}
x_1^{1}&=3,\quad y_2^{0}=4,\quad x_2^1=\frac{72}{7}\equiv \infty,
\quad y_2^1=\frac{14}{36}\equiv 0,\quad x_1^2=\frac{24}{7} \equiv \infty,\\
x_2^2&=\frac{120}{72} \equiv 4,\quad y_3^1=\frac{24}{9} \equiv 5.
\end{align*}
Thus all the values are uniquely determined over $\P\F_7$.
Figure \ref{figurekdv1} show a time evolution pattern of the discrete KdV equation \eqref{dKdV2} over $\P\F_7$ for the initial conditions $x_1^0=6,\ x_2^0=5,\ x_3^0=4,\ x_4^0=3,\  x_j^0=2\ (j\ge 5)$ and $y_1^t=2\ (t\ge 0)$.

This example suggests that the equation \eqref{dKdV2} should be understood as evolving over the field $\F_r(\delta)$, the rational function field with indeterminate $\delta$ over $\F_r$. 
To obtain the time evolution pattern over $\P\F_r$, we have to substitute $\delta$ with a suitable value $\delta_0 \in\F_r$ ($\delta_0=1$ in the example above).
This substitution can be expressed as the following reduction map:
\[
\F_r(\delta)^{\times}\rightarrow \P\F_r:\ (\delta-\delta_0)^s\frac{g(\delta-\delta_0)}{f(\delta-\delta_0)}\mapsto \left\{
\begin{array} {cl}
0  & (s>0),\vspace{1mm}\\
\infty & (s<0),\vspace{1mm}\\
\dfrac{g(0)}{f(0)} & (s=0),
\end{array}
\right.
\]
where $s\in \Z$, $f(h),g(h)\in\F_r[h]$ are co-prime polynomials and $f(0)\neq 0, g(0)\neq 0$.
With this prescription, we know that $0/0$ does not appear and we can uniquely determine the time evolution for generic initial conditions defined over $\F_r$.

\subsection{Soliton solutions of the discrete KdV equation over finite fields}

\begin{figure}
\centering
\includegraphics[width=12cm, bb=60 550 520 730]{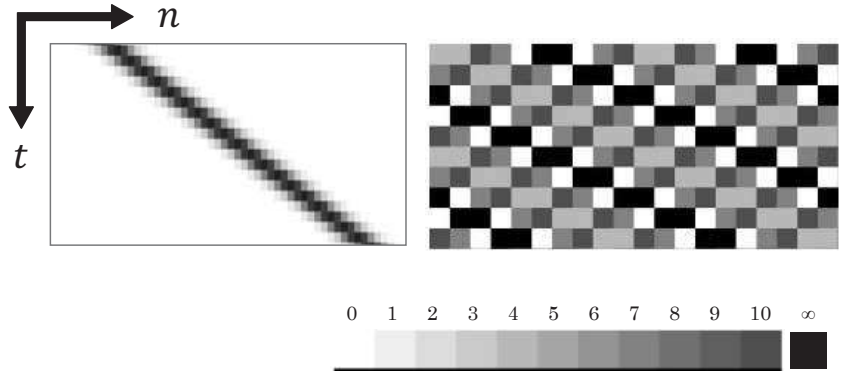}
\caption{The one-soliton solution of the discrete KdV equation \eqref{dKdV1} over $\R$ (left) and $\P\F_{11}$ (right) where $\delta=7,\ \gamma_1=2,\ l_1=9$. Elements of $\P\F_{11}$ are represented on the following grayscale: from $0$ (white) to $10$ (gray) and $\infty$ (black).}
\label{figure4}
\end{figure}

\begin{figure}
\centering
\includegraphics[width=12cm, bb=60 560 530 730]{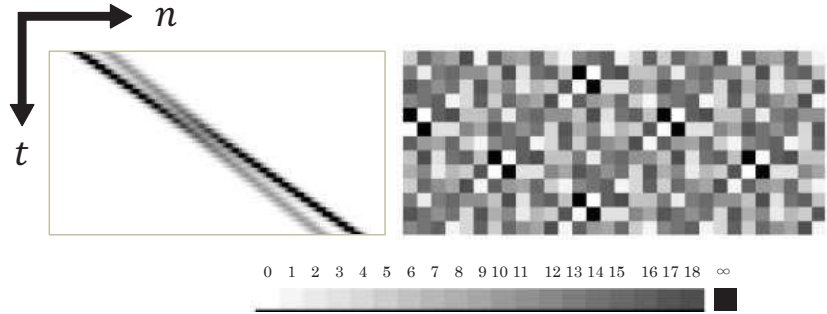}
\caption{The two-soliton solution of the discrete KdV equation \eqref{dKdV1} over $\R$ (left) and $\P\F_{19}$ (right) where $\delta=8,\ \gamma_1=15,\ l_1=2,\ \gamma_2=9,\ l_2=4$. Elements of $\P\F_{19}$ are represented on the following grayscale: from $0$ (white) to $18$ (gray) and $\infty$ (black). It is difficult to see the interaction of solitons over $\P\F_{19}$.}
\label{figure5}
\end{figure}

We consider the $N$-soliton solutions to \eqref{dKdV1} over $\F_r$.
As mentioned in the introduction, the $N$-soliton solution is given as
\begin{align*}
x_n^t&=\frac{\sigma_n^t\sigma_{n+1}^{t-1}}{\sigma_{n+1}^t\sigma_n^{t-1}},\\
\sigma_n^t&:=\det_{1\le i,j\le N}\left( \delta_{ij}+\frac{\gamma_i}{l_i+l_j-1}\left(\frac{1-l_i}{l_i}\right)^t
\left(\frac{l_i+\delta}{1+\delta-l_i}\right)^n\right)
\end{align*}
where $\gamma_i,\ l_i$ $(i=1,2,...,N)$ are arbitrary parameters but $l_i \ne l_j$ for $i \ne j$.
When $l_i,\ \gamma_i$ are chosen in $\F_r$, $x_n^t$ becomes a rational function in $\F_r(\delta)$. 
Hence we obtain soliton solutions over $\P\F_r$ by substituting $\delta$ with a value in $\F_r$.
The figures \ref{figure4} and \ref{figure5} show one and two soliton solutions for the discrete KdV equation \eqref{dKdV1} over the finite fields $\P\F_{11}$ and $\P\F_{19}$. The corresponding time evolutionary patterns on the field $\R$ are also presented for comparison.

The soliton solution of the discrete KdV equation over the finite field $\F_r$ has a period $r-1$ since we have $a^{r-1}\equiv 1$ for all $a\in\F_r^{\times}$.

\subsection{The discrete KdV equation over $\mathbb{Q}_p$}

We can also define the time evolution of the discrete KdV equation over the field $\F_p$ by considering the system over the field of $p$-adic numbers $\mathbb{Q}_p$ instead of $\mathbb{F}_p(\delta)$, in the same way as we have done to the discrete Painlev\'{e} equations.
We just have to compute the evolution of \eqref{dKdV2} or its soliton solutions over $\mathbb{Q}_p$ and then reduce the results to $\mathbb{F}_p$.
(To deal with the equation over $\F_{p^m}$ ($m\ge 2$), we require the field extension of $\Q_p$.)

\section{Concluding remarks}
In this article we investigated the discrete Painlev\'{e} equations and discrete KdV equation over finite fields.
To avoid indeterminacy, we examined two approaches. 
One is to extend the domain by blowing up at indeterminate points. 
According to the theory of the space of initial conditions, this approach is possible for all the discrete Painlev\'{e} equations.
An interesting point is that the space of initial conditions over a finite field can be reduced to a minimal domain because of the discrete topology of the finite field.
The other is to define the system over the larger field ($\mathbb{Q}_p$ in the case of dP\II, $q$P\I and $q$P\II, or $\mathbb{F}_p(\delta)$ in the case of discrete KdV equation), and then reduce the results to finite fields.
In particular the systems over $\mathbb{Q}_p$
attracted our attention.
We defined a new integrability test (\textit{almost good reduction}), which is an arithmetic analogue of the singularity confinement test, and proved that several types of the discrete Painlev\'{e} equations have this property.
Thanks to this property, not only the time evolution of the discrete Painlev\'{e} equations can be well defined, but also
a solution over $\Q$ (or $\Q_p$) can be directly transferred to a solution over $\P\F_p$.
This approach is equally valid in other discrete Painlev\'{e} equations and we expect that the same is true for its generalisations \cite{KNY}.
Furthermore, this `almost good reduction' criterion is expected to be applied to finding higher order \textit{integrable} mappings in arithmetic dynamics.
In the last section, we studied the discrete KdV equation and showed that a similar approach is also useful in defining the discrete partial difference equations such as soliton equations over finite fields and in obtaining the soliton solutions of these equations.
One of the future problems is to investigate the property of solitary waves of discrete KdV equation and other soliton equations such as discrete KP equation and discrete nonlinear Schr\"{o}dinger equation over the finite field. Solving the initial value problems for these equations over the field of $p$-adic numbers, the field of rational functions and the finite fields is also an important problem to be studied.

\section*{Acknowledgement}
The authors wish to thank Professors K. M. Tamizhmani and  R. Willox for helpful discussions.
This work is partially supported by Grant-in-Aid for JSPS Fellows (24-1379).

\end{document}